# Leveraging virtual technologies to enhance museums and art collections: insights from project CHANGES

Gianluca Genovese, gianluca.genovese@unisob.na.it, https://orcid.org/0000-0002-1064-2061, University of Suor Orsola Benincasa, Naples, Italy
Ivan Heibi, ivan.heibi2@unibo.it, https://orcid.org/0000-0001-5366-5194, University of Bologna, Bologna, Italy
Silvio Peroni, silvio.peroni@unibo.it, https://orcid.org/0000-0003-0530-4305, University of Bologna, Bologna, Italy
Sofia Pescarin, sofia.pescarin@cnr.it, https://orcid.org/0000-0002-9529-7083, Italian National Research Council, Florence, Italy

## Purpose

Our work is set in the context of Project CHANGES (https://www.fondazionechanges.org/), and in particular, within one of the research axes addressed in the project, which is dedicated to the use of virtual technologies for museums and art collections (Spoke 4 of the project). Several international policies support the focus on the universal use of Cultural Heritage (CH) digital data. Indeed, international, European, and national initiatives – guided, for instance, by UNESCO, the European Union, and the Italian Ministry of Culture – have provided specific guidelines for CH digitisation in the past years. To comply with these policies and guidelines, we need to make the digital enhancement of CH a permanent and widespread practice in CH institutions to increase the knowledge, curation and management of artefacts in all forms, expand the involvement of the general public, thus, improving accessibility, inclusiveness, critical thinking, participation, enjoyment and sustainability.

The work conducted in Spoke 4 wants to provide tools and prototypes to address such a digital enhancement in the context of museums and art collections. In particular, it focuses on answering the following research questions:
1. How do we preserve and make accessible CH, specifically the potential experience a visitor can have with it?
2. How can environmental, temporal, and contextual factors impact the digital result of digitisation processes?
3. Which material, methods, and tools can be used to implement a reproducible workflow for acquiring and digitising 3D cultural heritage artefacts?

# Methodology

To address these questions, we have experimented with different *templates* of museums and art collections for designing case studies and best practices that can be further adapted and reused in institutions and contexts sharing similar characteristics. In this context, we have considered a set of scenarios which involve heterogeneous cultural institutions, which include natural history and scientific museums, widespread art galleries, site museums with (in)tangible heritage and landscapes, historical palaces, demo-ethnic anthropological museums, and museums with extensive collections and high-tech approaches.

We have identified a set of cultural institutions with which to work compliantly with the templates mentioned above. We have set up specific research projects (i.e. nine "core" case studies) with them to put into practice existing and ad-hoc tools developed in the project to provide, in perspectives, a system to be freely reusable in a similar context by other cultural institutions not necessarily involved in the project. These institutions are the Egyptian Museum in Turin, University of Bologna's Museum System – Geological Collection "Giovanni Capellini Museum", University of Ferrara's Museum System, University of Turin's Museum System, Ente morale Istituto Suor Orsola Benincasa in Naples, Centro Studi e Archivio della Comunicazione (CSAC) in Parma, Carlo Levi Literary Park in Aliano, Grazia Deledda Literary Park in Galtellì, The Royal Palace of Caserta.

Before working with these case studies, we run a pilot study to gather guidelines for digitising museum and art collections. As a pilot, we have identified a scenario that could serve as a common experimental ground: a temporary exhibition (ended on May 28, 2023) containing a large and heterogeneous set of different small/medium objects to be acquired and entitled "The Other Renaissance: Ulisse Aldrovandi and the Wonders of the World" (Balzani et al., 2024). Within this pilot, we experimented the adoption and development of a plethora of virtual (energy efficient) technologies: decentralised and interlinked knowledge graphs of digital CH and (in)tangible objects, Web-based environments for sharing cultural heritage and involving users in museums and art collections in situ or remotely, new design approaches to virtual technologies for cultural heritage including eXtended Reality (XR, i.e. virtual reality, augmented reality, mixed reality, immersive reality), gamification, serious games, edutainment, 2D/3D models and multimedia (including video storytelling), tools for digitisation and simulation for enabling digital approaches to cultural heritage, Internet of Things and sensors networks, AI-based methods and tools for cultural heritage, location-based technologies connected to GIS for cultural heritage (e.g. sites), lighting technology for exhibitions.

The guidelines produced (Pescarin et al., 2024) have been shared with all the partners working in the nine "core" case studies to guide the digitisation process. This has been conducted in collaboration with companies providing implementation efforts and institutions providing domain cultural knowledge and citizen engagement. The companies involved were selected via a public call for applications supported by the project's funds.

# Findings

The project allowed us to define suitable research lines with the CH institutions involved, which also served with next year's plans for such institutions, aiming at mutual-beneficial collaboration. All the necessary methodologies and tools to implement such research lines have been tested first in the creation of a digital twin of a temporary exhibition dedicated to Ulisse Aldrovandi, entitled to investigate the use of virtual technologies for the promotion, preservation, exploitation and enhancement of cultural heritage in museums and art collections (Barzaghi et al., 2024b). One of the outcomes of this experimentation has been the identification of a workflow, methodologies and tools to produce cultural heritage data (datasets, software, 2D and 3D models, etc.) compliant with the FAIR principles and following strictly Open Science principles (Barzaghi et al., 2024a).

All these research outcomes have been used systematically to implement the research of the nine "core" case studies and answer our three research questions. These "core" case studies have also been accompanied by additional research studies aligned with the project templates and goals, which the partners of Spoke 4 have used to experiment with further adoptions of the aforementioned workflow, methodologies and tools within the project's period. In addition, we have identified potential stakeholders interested in the outcomes of the "core" case studies (Genovese et al., 2024) and have implemented tools to monitor and assess both the "core" case studies and the involvement and exploitation of all the research products and expertise created during the project.

# Value

The overall goal of Spoke 4 was twofold. On the one hand, we wanted to experiment using the aforementioned technologies in defined and representative real-case scenarios. On the other hand, the work to handle in the context of these CH institutions has allowed us to define and share guidelines for instructing institutions and researchers on the processes and needs to set up appropriate workflows to acquire and digitise CH in a way that is compliant with European policies on Open Science, and to develop a set of open tools and data that can also be reused by other national, European, and international institutions that were not involved directly in the research of the Spoke, to maximise the investments of Project CHANGES and to enable future users to adopt these technologies after the project finish.

Although Spoke 4's outcomes were developed with the primary goal of enhancing CH, cross-fertilisation between different disciplines and unexpected applications in additional settings can arise. This scenario can be feasibly addressed only if the methodologies and tools developed, as in our case, permit much flexibility to be reused in different contexts. For instance, we have recently adapted the first two rooms of the digital twin implemented within our pilot study for work with psychologists, designers, and digital heritage experts to analyse user behaviour in 3D and VR environments (Massidda et al. 2024). Another possible application is to reuse the digital CH objects produced during the acquisition and digitisation campaigns organised within Spoke 4 activities to create new publishing and editorial artefacts that may serve, for instance, to develop innovative teaching sessions with students. For example, using the Smithsonian Voyager tool (https://smithsonian.github.io/dpo-voyager/, an open-source 3D explorer and authoring tool suite), a digital heritage expert and a professor expert in Mesoamerican pictorial manuscripts and mosaics have created an interactive scholarly digital edition (shown in Figure 1, https://pure3d.eu/new-3d-scholarly-edition-codex-cospi), with different narrative dimensions explorable by the user, of the Cospi Codex (https://bub.unibo.it/it/bub-digitale/il-codice-cospi/bub_ms4093_cospi), one of the 3D pieces digitised from the Aldrovandi's temporary exhibition (i.e. our pilot study), which is one of only thirteen precolonial Mesoamerican manuscripts globally available created by Eastern Nahua painters between the 15th and early 16th centuries. Such a scholarly digital edition will be used in the context of a PhD programme to present to students both the research around Mesoamerican studies and the possibilities that digital tools and 3D modelling can offer for providing a new way of teaching Humanities research.

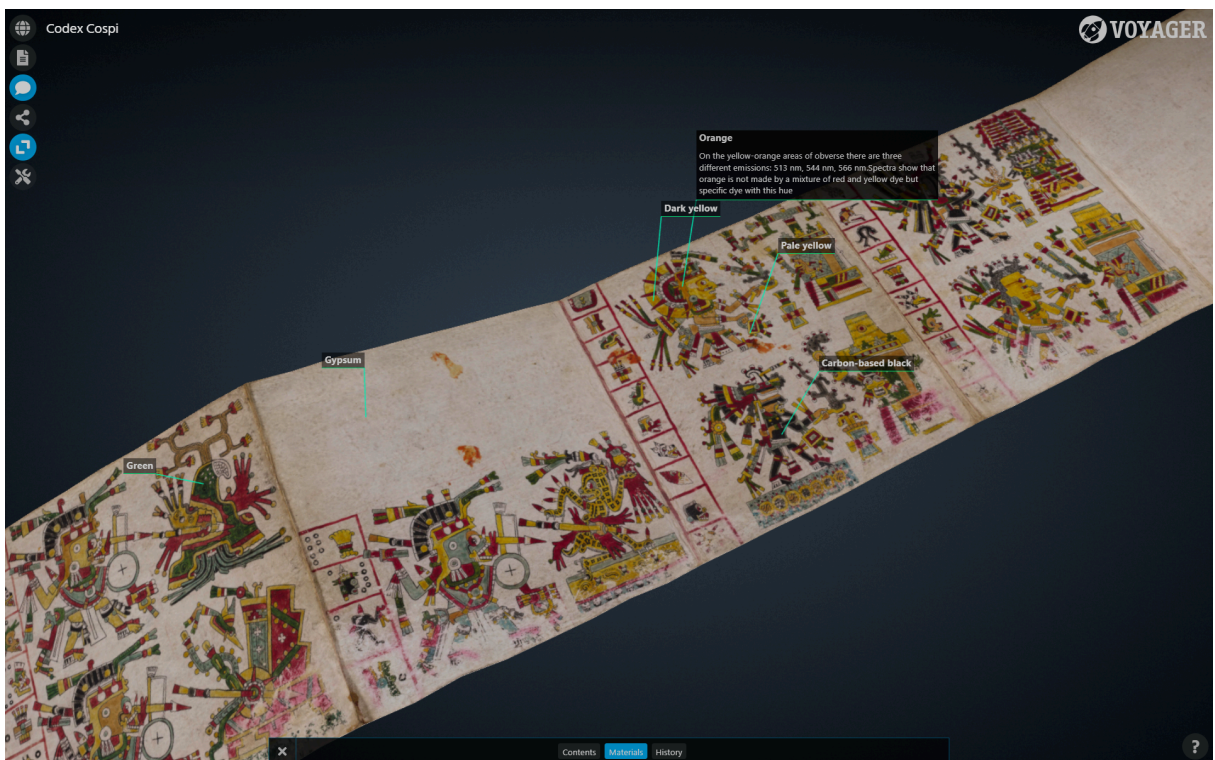

**Figure 1.** The scholarly digital edition of the Codex Cospi in the Smithsonian Voyager tool.

# Acknowledgements

This work has been funded by Project PE 0000020 CHANGES - CUP B53C22003780006, NRP Mission 4 Component 2 Investment 1.3, Funded by the European Union - NextGenerationEU.